# Systematic Review on Healthcare Systems Engineering utilizing ChatGPT

**Jungwoo Kim**[1]　　**Ji-Su Lee**[2]　　**Huijae Kim**[1]　　**Taesik Lee**[1,2,†]

KAIST　　KAIST　　KAIST　　KAIST

jungwoo415@kaist.ac.kr　　jisulee@kaist.ac.kr　　huijae.kim@kaist.ac.kr　　taesik.lee@kaist.edu

[1]Department of Industrial and Systems Engineering, KAIST, Daejeon, South Korea

[2]Graduate School of Data Science, KAIST, Daejeon, South Korea

This paper presents an analytical framework for conducting academic reviews in the field of Healthcare Systems Engineering, employing ChatGPT, a state-of-the-art tool among recent language models. We utilized 9,809 abstract paragraphs from conference presentations to systematically review the field. The framework comprises distinct analytical processes, each employing tailored prompts and the systematic use of the ChatGPT API. Through this framework, we organized the target field into 11 topic categories and conducted a comprehensive analysis covering quantitative yearly trends and detailed sub-categories. This effort explores the potential for leveraging ChatGPT to alleviate the burden of academic reviews. Furthermore, it provides valuable insights into the dynamic landscape of Healthcare Systems Engineering research.

## 1. Introduction

Academic review is a process of producing a tourist guidebook, providing researchers a practical guidance through the landscape of a certain field. They offer well-digested summaries of contemporary literature, standards of approaches, and evaluations thereof (Erren, et al., 2009). Moreover, they provide dynamic trends of research interest, including the sites where new knowledge is emerging and which achievements are needed in the future (Ketcham, et al., 2007). However, performing an academic review requires extensive reading and in-depth knowledge of the up-to-date academic archives of the field. Consequently, these reviews heavily rely on the invaluable knowledge and timely contributions of veteran scholars. (Pautasso, 2013)

In this project, we explored the opportunity to reduce the burden of such a task, by leveraging ChatGPT. As a cutting-edge tool among recent language models, ChatGPT allows input prompts to be in highly versatile formats, including human-spoken sentences. With this capability, we were able to build the entire framework of an academic review, by weaving together distinct analysis processes as components. Most of these processes were implemented using tailored input prompts, and systematic utilization of ChatGPT API. This framework significantly reduced the human and time resources required, compared to the conventional handmade academic reviewing process. It took less than three weeks, to read and organize approximately 40,000 research abstracts, and even going through all the trials and errors during the structurization of the analysis framework,

The target field of our analysis is **Healthcare Systems Engineering**, consisting of the applications of systems engineering methodologies to address various challenges within Healthcare Systems. We analyzed a seven-years-collection of research abstracts presented at the relevant conferences: INFORMS Annual Meetings and INFORMS Healthcare Conferences. The product of our review framework consists of 11 major topic categories, along with detailed expositions of the inner contents and yearly trends regarding each.

In addition to introducing the academic review framework using ChatGPT, the parallel goal of this work is to provide the academic review itself, of the Healthcare Systems Engineering. The dominant layout of the research within this field includes model formulation, analysis, and solving the problems within the system, rather than advancing the generalized engineering techniques themselves. Therefore, our intention was more inclined towards how researchers address problems from various domain systems of healthcare, which is thought to be particularly valuable for newcomers to this field.

Chapter 2 provides some descriptions of relevant literature, with a particular focus on efforts utilizing ChatGPT and other language models for literature reviews. In Chapter 3, we present an overview of the entire review framework, including detailed specifications of the dataset and the ChatGPT models employed. Chapters 4 to 6 cover the detailed processes and outcomes of the three major steps in our analysis: Dataset Preparation, Topic Categorization, and Comprehensive Analysis. Chapter 7 provides some discussions with two perspectives of our work: the contents of our academic review, and the analysis framework we built. This section offers in-depth interpretations and opinions on the results of our analysis and also provides insights and comments on the utilization of ChatGPT into the literature review task. In conclusion, Chapter 8 summarizes the accomplishments of our work.

## 2. Related Works

The emergence of ChatGPT has led many researchers to employ it in their research processes since ChatGPT is capable of not only question and answering and text generation but also text classification, inference, and information extraction. (Liu, et al., 2023) For example, many recent studies have shown that GPT can be used to generate realistic data by harnessing its generative capabilities (Jansen, et al., 2023, Joon, et al., 2023) and analyzing data with the assistance of prompts (Maddigan, et al., 2023, Sharma, et al., 2022).

Furthermore, ChatGPT can be applied in literature review with the procedure of finding relevant research papers, understanding them, and categorizing them properly into several topics or methodologies. (Huang, et al., 2023) For instance, one study found that ChatGPT has the capability in generating effective Boolean queries for conducting systematic review literature searches and leads to rapid reviews by reducing the time spent searching for the search keywords to find suitable papers (Wang, et al., 2023) Besides the field of systematic review, several studies have suggested that ChatGPT can be applied in information extraction(Wei, et al., 2023), abstractive summarization(Soni, et al., 2023, Liu, et al., 2023) and text classification(Clavie, et al., 2023, Kuzman, et al., 2023) into some categories through its strength of availability in paucity or absence of data.

However, little attention has been paid to employing ChatGPT in the entire process of literature review. A recent study focused on automating the systematic review procedure using a practical implementation. The study utilized ChatGPT to generate relevant keywords and phrases to search papers, filtering studies included in the previous step, and classifying them into several subcategories. (Alshami, et al., 2023) However, the framework of systematic review proposed in the research assumed that reviewers are experts in the field of a topic that they are willing to review since they should derive subcategories of the topic by themselves to let ChatGPT categorize studies into them.

Most studies on the topic of literature review assisted by ChatGPT have only focused on specific parts of the process of review, thus researchers' expertise is necessary to review a lot of papers. However, this paper attempts to show that ChatGPT can write a structuralized review paper based on its reasoning and generating ability. We devise a procedure of review to utilize ChatGPT with suitable prompts in the overall process of review. As a result, we find that it can identify the basis of a topic from the relevant papers and give an output of the structure of them.

## 3. Review Procedure

### 3.1 Overview of the Analysis Framework

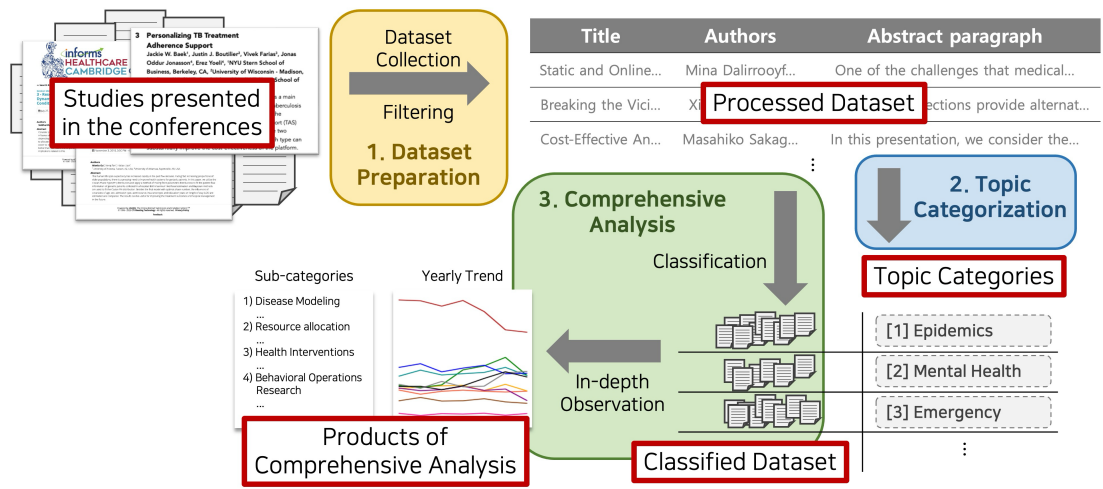

Figure 1. Overview of the Review Procedure with ChatGPT

The analysis framework is composed of four pivotal steps: Dataset Preparation, Topic Categorization, and Comprehensive Analysis consisting of two dimensions (Quantitative Analysis, and Qualitative Analysis). Dataset Preparation starts with the collection of the data, including the titles and abstracts of relevant studies presented at designated conferences. Then, the entries out of the scope of our analysis (those irrelevant to healthcare systems), were systematically filtered out using ChatGPT. Moving to the Topic Categorization process, we derived the topic categories that best describe the problem domain of the research in the dataset. This step serves as the construction of the backbone of our review and is also thought to be the keystone of our work.

With the topic categories derived, Comprehensive Analyses were conducted to explore deeper into each of them. Each study in the dataset was classified into the topic categories established previously, to gather the research that aligns with the scope of each category. This enables a quantitative measurement of their distribution as well, thus facilitating the observation of volumes and yearly trends of the research community's focus of interest over time. Furthermore, in-depth derivation of detailed topics with higher resolution was conducted within each category. This investigation identified the sub-categories, enhancing the understanding of more detailed examples of research problems.

### 3.2 Data Specification

The scope of our analysis covers the studies presented at two conferences that are relevant to our area of interest, within the INFORMS (Institute for Operations Research and the Management Sciences) community: the "INFORMS Annual Meeting" and the "INFORMS Healthcare Conference." INFORMS is a prominent academic community that spans the fields of Operations Research, Management Science, Data Science, Economics, and other related areas. Within this community, several conferences are held, with the INFORMS Annual Meeting standing out as the major event, featuring around 5,000 studies presented annually. The INFORMS Healthcare Conference is an event of a more specialized community, focused on the healthcare domain, with approximately 500 studies presented every two years.

The conferences incorporate various forms of academic products, including technical sessions, posters, tutorials, workshops, and plenary sessions. Researchers often use these platforms to report their progress and share some findings. While conventional academic reviews typically focus on officially published manuscripts, we decided to examine research presented at the conferences to better capture the diverse and experimental approaches. We believed that these somewhat unfiltered reports would provide a more comprehensive and up-to-date view of the community's activities.

### 3.3 Model Specification of ChatGPT

The ChatGPT models utilized in our review process were 'gpt-3.5-turbo' and 'gpt-4'. The maximum limit for input prompts for these models was set at 4,000 tokens for 'gpt-3.5-turbo' and 8,000 tokens for 'gpt-4' at the time of our usage. The API pricing for the 'gpt-3.5-turbo' model was $0.0010 per 1,000 tokens input and $0.0020 per 1,000 tokens output. For the 'gpt-4' model, the API pricing was $0.03 per 1,000 tokens input and $0.06 per 1,000 tokens output.

While the 'gpt-4' model boasts superior performance compared to the 'gpt-3.5-turbo,' it comes at a higher cost. exclusively for the production of the main topic categories and writing comprehensive explanations for them. The entire execution of our work incurred a total cost of less than $200, including all the trial-and-error phases throughout our endeavor.

## 4. Dataset Preparation

### 4.1 Data Collection

The data for the analysis was systematically gathered from nine conferences, including six Annual Meetings spanning from 2017 to 2022 and three Healthcare Conferences held in 2019, 2021, and 2023. Notably, although the 2023 Annual Conference took place from October 15 to 18, 2023, our data collection process was completed before this event. Our data acquisition primarily involved web-based abstract libraries associated with INFORMS Conferences (cOASIS, 2017; cOASIS, 2018; cOASIS, 2019; cOASIS, 2020; cOASIS, 2021; cOASIS, 2022; cOASIS, 2019; cOASIS, 2021; cOASIS, 2023). Leveraging a web-crawling tool developed using Selenium and Python, we systematically extracted crucial components from each presentation, including the Title, Authors, and Abstract paragraphs.

Table 3. Number of collected presentations in each conference

| Year | INFORMS Annual Meeting | | INFORMS Healthcare Conference |
|---|---|---|---|
| | Number of Studies | Number of Healthcare-related Studies (filtered) | Number of Healthcare-related Studies |
| 2017 | 5,234 | 1,316 (25.14%) | - |
| 2018 | 5,340 | 1,411 (26.42%) | - |
| 2019 | 6,797 | 1,734 (25.51%) | 512 |
| 2020 | 4,819 | 1,226 (25.44%) | - |
| 2021 | 6,510 | 1,415 (21.71%) | 410 |
| 2022 | 4,867 | 1,279 (26.28%) | - |
| 2023 | - | - | 506 |

Table 1 provides a detailed overview of the amount of collected data for each conference. The Annual Meetings contributed to a substantial portion, with an aggregate of approximately 5,500 studies, while each Healthcare Conference added around 500 studies. Consequently, the cumulative number of studies gathered was 34,995, and this number includes those that are beyond the scope of healthcare system problems.

## 4.2 Filtering

To filter out the studies unrelated to the healthcare system, we employed a systematic filtering process using ChatGPT. This filtering was specifically applied to the INFORMS Annual Meeting dataset, as the INFORMS Annual Meeting covers a range of industries, including Transportation, Manufacturing, Logistics, and more. In contrast, the INFORMS Healthcare Conference is a domain-specific event that aligns closely with our target field of analysis.

The ChatGPT prompt for the filtering process provided the information on each study, including the title and abstract paragraph, and asked ChatGPT whether the study was healthcare-system-related or not. The detailed prompt can be found in Appendix 1. The prompt was designed to limit responses in the format of 'Yes' or 'No.' In cases where the response deviates from these two options, ChatGPT was iteratively asked again until the desired response format was obtained.

The results of the filtering process are presented in Table 1. Approximately 1,300 healthcare-system-related studies were identified each year, constituting around 24% of the entire studies in the Annual meetings. There was no significant upward or downward trend in this proportion, and it remained above 25% except in 2021. The number of healthcare-related presentations in our entire dataset amounted to 9,809, which was utilized for the remaining steps of our review procedure.

The reliability of this filtering process was assessed using human-labeled data, consisting of 20 healthcare-system-related studies and 20 non-healthcare studies. The healthcare-related studies were randomly sampled from the INFORMS Healthcare Conference dataset, and the non-healthcare studies were selected from the INFORMS Annual Meeting dataset, with the manual exclusion of healthcare-system-related studies. ChatGPT successfully filtered 20 non-healthcare studies from the 40 samples.

The consistency of the filtering process was also evaluated by repeating the same task on a subset of the dataset. One hundred studies were randomly sampled from the INFORMS Annual Meeting dataset, and the filtering process was repeated five times for these samples. Only two out of the 100 studies had non-uniform outputs, both having 4:1 votes (4 yes, 1 no / 1 no, 4 yes). We have concluded that the reliability and consistency of this filtering process utilizing ChatGPT were sufficient for exploring this research area.

## 5. Topic Categorization

Topic categorization serves as the cornerstone of our analysis framework. The primary objective is to organize the extensive array of studies in the field by categorizing them based on their target problem domain or system. Through this process, we have derived 11 topic categories that best describe the research endeavors in this field.

### 5.1 Keyword Extraction

To execute this process, we attempted to provide ChatGPT with numerous research studies and asked it to categorize them. However, a challenge arose due to the input token limit of the gpt-3.5 model, which was 4,000 tokens at the time of our usage, while the average length of abstracts in our database was 145.2 tokens. Although some models allowed for longer input, the model with the maximum length had a limit of 16,000 tokens, accommodating only an average of 110 researchers.

As a solution, we conducted a Keyword Extraction process to condense the input prompt for the Topic Categorization process. This process aimed to eliminate stop words and non-essential expressions to capture the essence of each study. The detailed prompt for the Keyword Extraction process can be found in Appendix 2. The prompt requested 5 keywords describing the problem domain of each study and 5 keywords about the technical methodologies employed. This approach aimed to narrow down the derived categories to address similar elements in each study. However, we excluded the analysis of the technical methodologies, as meaningful insights could not be derived. The length of each paragraph was condensed from an average of 145.2 tokens to 5 problem domain keywords with an average of 8.9 tokens.

### 5.2 Categorizing

The Topic Categorization process was executed utilizing the list of extracted keywords from the paragraphs, Since this task is considered to be the core part, and demands the highest level of linguistic proficiency in our analysis procedure, the 'gpt-4' model was used only for this task considering expensive API pricing.

Because the entire list of keywords didn't fit the token count limit of the input prompt, 2500 keywords were randomly sampled to consist of a comma-separated list of keywords, The detailed prompt of Topic Categorization process can be found in Appendix 3. The topic categorization was conducted 10 times with all different sampled lists of keywords.

## 5.3 Curated Topic Categories

10 trials of topic classification results had quite similar outputs, with approximately 10 topic categories produced. However, some similar categories were merged and separated in each trial, and also the expression for the same concept was slightly different for each case. 3 instances of these raw outputs of categories are presented in Appendix 4. To break through this ambiguity the result, we (the authors, who are human researchers) have curated and finalized the version of topic categories based on ChatGPT's result and the researcher's expertise. Following is the list of curated 11 topic categories.

Table 4. Curated Topic Categories of Healthcare Systems Engineering

| Topic Category 1 : Clinical Decision Making |
|---|
| This domain focuses on decision-making strategies and disease management in patient treatment. It includes research on randomized experiments, individual treatment effects, and screening strategies related to patient discharge. There is a specific emphasis on managing various diseases such as cardiovascular disease, chronic kidney disease, cancers, and ALS(Amyotrophic Lateral Sclerosis), with research areas covering disease progression management, treatment development, and evolving disease trends. |
| Topic Category 2 : Patient Care Management |
| This category concentrates on care methods that directly affect patient health, including shared medical appointments, patient engagement, personalized care, home care, patient preferences and choices, and risk prediction algorithms. |
| Topic Category 3 : Healthcare Resource Management |
| This category explores clinical practices and efficient healthcare resource allocation. It includes topics such as treatment plans, preventive care, outpatient appointment scheduling, and reducing patient waiting times and no-shows. It also covers efficient resource allocation in healthcare, including medical supplies, facilities, and human workforce. The goal is to enhance and streamline the hospital operations and bring about improvements in patient care and satisfaction. |
| Topic Category 4 : Public Health and Epidemiology |
| Topics here address disease management on a larger scale, such as population health or epidemic diseases. These researches address diseases like Hepatitis C, |

Malaria, AIDS, COVID-19, or Monkeypox. They utilize spatial dynamics and simulation tools to express disease transmission, and invent preventive treatments or interventions and policies. It also includes research about social impacts of public health rules such as opioid crisis.

**Topic Category 5 : Health Policy and Economics**

This domain examines the impact of payment systems and policies such as Medicare, Medicaid, Fee-for-Service, and Accountable Care Organizations on the healthcare system. It also includes studies on forecasting health systems during unforeseen medical circumstances, healthcare economics, and policy. Topics like subsidies for rare disease therapeutics, public vs. private hospitals, and vaccination affordability are explored. It delves into how economic factors and government policies affect population health and healthcare service delivery.

**Topic Category 6 : Preventive Healthcare and Well-Being**

This category includes research on preventive treatments or behaviors effecting health. The impact of lifestyles such as physical activity, subsidy programs, and diet on health outcomes, and the deployment of preventive services like vaccines are studied.

**Topic Category 7 : Emergency and Critical Care**

Here, topics encompass emergency departments, critical incidents, disasters, emergency medical services, and intensive care units. These areas involve immediate and life-saving actions and decisions.

**Topic Category 8 : Psychiatric Care and Mental Health**

This category involves the study of mental health diseases and disorders, treatment courses, early identification through healthcare data, and insurance claims. It also covers aspects of mental health and psychiatry, including psychiatric readmissions, substance disorders, and mental healthcare access.

**Topic Category 9 : Innovations in Healthcare**

Topics related to innovative healthcare approaches, the implementation of new technologies and services, and digital transformations in healthcare, including telemedicine, E-consult and online patient portals, fall under this category.

**Topic Category 10 : Health Informatics an AI**

Research in this category involves applying machine learning and data science to improve patient care, manage hospital resources, and predict health outcomes. This includes the use of electronic health records, big data analysis, artificial intelligence, crowd intelligence, and wearable devices.

**Topic Category 11 : Social Determinants of Health and Health Inequalities**

> This category focuses on understanding how social, economic, and environmental factors impact health outcomes, especially for underserved and marginalized communities. Research areas include racial and socioeconomic disparities, and the impact of social determinants on diseases.

The categorization from 1 to 5 is based on different levels of decision-making and management. Category 1 predominantly focuses on therapeutic decision-making for individual patients at the micro level. It encompasses research aimed at supporting decision-making at the micro level, not only in therapeutic decisions but also in diagnoses and disease management. Consequently, the title has been modified to "Clinical Decision Making." While Category 1 primarily addresses therapeutic decision-making and management, Category 2 is centered around non-therapeutic decision-making and management, leading to the exclusion of "quality" from the title, considering its less explicit or specific relevance to healthcare quality.

Category 3 emphasizes research at the meso level, focusing on institutional-level management and decision-making. Categories 4 and 5 seem to pertain to decision-making and management at the macro level. The primary focus here lies in influencing populations at large, with an emphasis on infectious diseases, policies, and other macro-level considerations.

Moving to Categories 6 to 8, these represent specialized areas of study. Category 6 delves into preventive interventions and well-being, while Category 7 focuses on emergency and critical care. Category 8 addresses psychiatric treatment and mental health.

Categories 9 and 10 primarily address technological advancements. While there is a slightly ambiguous overlap between the two, Category 9 focuses more on the innovation in the Healthcare service itself, while Category 10 concentrates more on applying up-to-date information technology, data science, and artificial intelligence to the healthcare systems.

Finally, Category 11 stands apart from the preceding research topics, addressing social determinants of health and health inequalities.

## 6. Comprehensive Analysis

To produce some further comprehensive insights on the above 11 topic categories, two types of in-depth analysis were conducted: Quantitative and Qualitative analysis.

The purpose of the Quantitative analysis is to measure the level of academic community's interest associated with each category, and the temporal changes of this interest. For the Qualitative analysis, detailed sub-topics distributed within each category were further derived. This was done to figure out the mainstream approaches or challenges of them. This section provides methodological details and concise results of the in-depth analysis, with further discussions on the results presented in the subsequent Discussion section.

Table 5. Volume and Classified studies in each Topic Category and Year

| Year | | | | | | | Total |
|---|---|---|---|---|---|---|---|
| 2017 | 2018 | 2019 | 2020 | 2021 | 2022 | 2023 | |
| Topic Category 1 : Clinical Decision Making | | | | | | | |
| 218 (16.6%) | 203 (14.4%) | 358 (15.9%) | 203 (16.6%) | 291 (15.9%) | 199 (15.6%) | 81 (16.0%) | 1553 (15.8%) |
| Topic Category 2 : Patient Care Management | | | | | | | |
| 230 (17.5%) | 257 (18.2%) | 424 (18.9%) | 235 (19.2%) | 299 (16.4%) | 252 (19.7%) | 82 (16.2%) | 1779 (18.1%) |
| Topic Category 3 : Healthcare Resource Management | | | | | | | |
| 858 (65.2%) | 914 (64.8%) | 1390 (61.9%) | 796 (64.9%) | 1076 (59.0%) | 628 (49.1%) | 242 (47.8%) | 5904 (60.2%) |
| Topic Category 4 : Public Health and Epidemiology | | | | | | | |
| 251 (19.1%) | 254 (18.0%) | 436 (19.4%) | 337 (27.5%) | 541 (29.6%) | 438 (34.2%) | 125 (24.7%) | 2382 (24.3%) |
| Topic Category 5 : Health Policy and Economics | | | | | | | |
| 318 (24.2%) | 386 (27.4%) | 556 (24.8%) | 312 (25.4%) | 477 (26.1%) | 369 (28.9%) | 124 (24.5%) | 2542 (25.9%) |
| Topic Category 6 : Preventive Healthcare and Well-Being | | | | | | | |
| 380 (28.9%) | 435 (30.8%) | 594 (26.4%) | 349 (28.5%) | 547 (30.0%) | 325 (25.4%) | 130 (25.7%) | 2760 (28.1%) |
| Topic Category 7 : Emergency and Critical Care | | | | | | | |
| 232 (17.6%) | 230 (16.3%) | 378 (16.8%) | 229 (18.7%) | 300 (16.4%) | 220 (17.2%) | 56 (11.1%) | 1645 (16.8%) |
| Topic Category 8 : Psychiatric Care and Mental Health | | | | | | | |
| 53 (4.0%) | 41 (2.9%) | 88 (3.9%) | 45 (3.7%) | 75 (4.1%) | 39 (3.0%) | 20 (4.0%) | 361 (3.7%) |
| Topic Category 9 : Innovations in Healthcare | | | | | | | |
| 143 (10.9%) | 178 (12.6%) | 240 (10.7%) | 133 (10.8%) | 218 (11.9%) | 131 (10.2%) | 48 (9.5%) | 1091 (11.1%) |
| Topic Category 10 : Health Informatics an AI | | | | | | | |
| 259 (19.7%) | 248 (17.6%) | 478 (21.3%) | 225 (18.4%) | 342 (18.7%) | 331 (25.9%) | 135 (26.7%) | 2018 (20.6%) |
| Topic Category 11 : Social Determinants of Health and Health Inequalities | | | | | | | |
| 244 (18.5%) | 263 (18.6%) | 378 (16.8%) | 232 (18.9%) | 418 (22.9%) | 339 (26.5%) | 120 (23.7%) | 1994 (20.3%) |

6.1 Classification process

To facilitate these in-depth analyses, it was necessary to gather all the studies that belong to each category. Thus the Classification process for all 9,809 studies, assigning each to its respective categories, was executed. The detailed prompt of the Classification process can be found in Appendix 5. The prompt allowed ChatGPT to assign the studies into multiple topic categories, as well as one, or even none of them. But there was no instance that didn't belong to any of the categories, indicating that the topic categories cover the area pretty well. We also asked for a one-sentence summary of each study, for the use of sub-topic derivation.

Since ChatGPT is a generative model, the output of this Classification process may differ for each trial. A hard-voting system was implemented to handle this stochasticity. We repeated the process 5 times each for the entire dataset and assigned it to the category according to hard voting with criteria of 4 votes. In other words, the research is assigned to a certain category if more than 4 out of 5 answers say that the research belongs to it. The studies that didn't get more than 4 votes for any of the categories were manually classified.

### 6.2 Quantitative Analysis: Yearly trend

Table 3 shows the number of studies classified to each topic category for each year, with the proportions in the bracket. The total number and proportion of studies exhibit a relatively uniform proportion ranging from 10% to 30%, except for categories 3 and 8. Category 3 stands out with an overwhelming 60.2% proportion, indicating a substantial amount of interest in meso-level healthcare systems, specifically at the level of healthcare institutions. In contrast, category 8 has a minimal proportion of 3.7%, indicating relatively less volume of system engineering approaches with mental health.

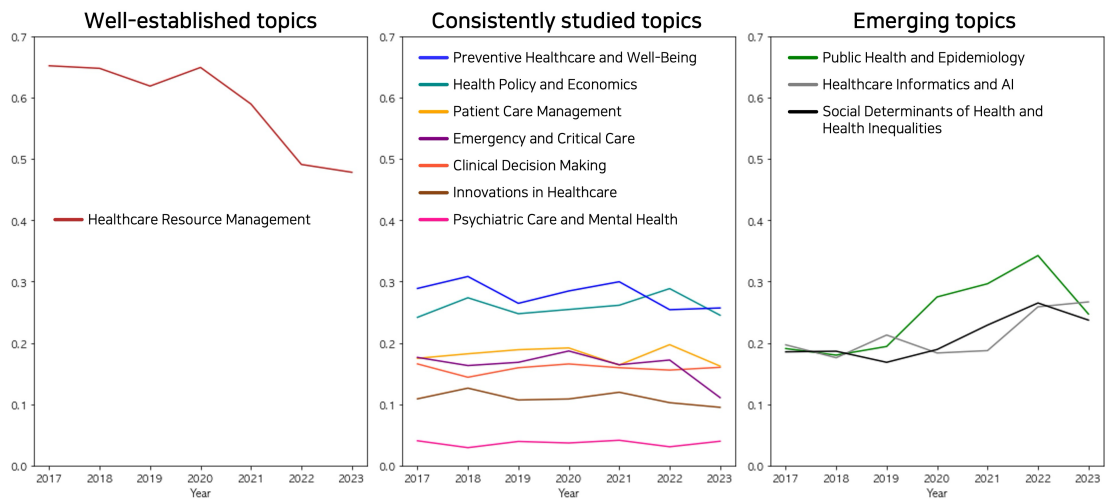

Figure 2. Overview of the Review Procedure with ChatGPT

The topic categories can be labeled into three types according to their yearly trend: **Consistently studied topics** maintaining a stable proportion throughout the study period, **Well-established topics** experiencing a decline in recent research compared to the past, and **Emerging topics** witnessing an increase in recent conferences. The labeling criterion was based on simple linear regression lines fitted to the yearly proportions of each category, with those having a slope exceeding 1% annually labeled as Emerging, while those below -1% were labeled as Well-established. Figure 2 illustrates the plot of each category's proportions in the field, organized with their types mentioned.

### 6.3 Qualitative Analysis : Sub-topics

We further delved into understanding the content of each topic category, by inspecting the sub-topics within them. Sub-topics are more specified topics or problem domains that form a mainstream in a certain topic category. For example, 'Telemedicine' is one of the sub-topics in the topic category 'Innovations in Healthcare'. One-sentence-summaries produced in the Classification process were utilized for this since the list of extracted keywords somewhat lacked the amount of details. To match the token limit of the input prompt, a maximum of 500 one-sentence-summaries were sampled from those researches that belong to each topic category. The detailed prompt for the Sub-topic mining process can be found in Appendix 6. The list of sub-topics derived for each topic category can be examined through Appendix 7, and reviewed in Discussion.

### 7. Discussion

## 7.1 Discussion on the review contents

This section encompasses discussions focusing on the contents of our review, which include the Volume of the data (Table 1), Curated Topic Categories (Table 2), Results of Quantitative Analysis (Table 3 and Figure 2), and Sub-topics (Appendix 7). Through reflecting on research topics and trends within the field, our primary aim is to deliberate directions for advancement and progress in Korea's Healthcare Systems Engineering research efforts moving forward. Our analysis was on the major conferences in the United States, which have substantial academic legacies in Engineering-based research within the Healthcare System field. The substantial variations in healthcare system characteristics among different societies and nations should be noted. Consequently, research endeavors in this domain should be tailored to the specific characteristics of the country and society. Hence, the insights provided by the analysis could offer guidance for the academic trajectory of Korea, thus producing achievements suitable for Korean Healthcare Systems.

As illustrated in Table 1, the INFORMS Annual Conference showcases a notable proportion of studies, ranging from 20-25%, corporating Healthcare systems. This percentage is particularly noteworthy considering the conference's comprehensive coverage spanning domains such as transportation, logistics, and manufacturing, in addition to healthcare. Such a statistic emphasizes a significant scholarly interest in addressing systemic issues within healthcare systems. It also emphasizes the applicability of methodologies drawn from Industrial Engineering and Management Science to this domain.

Examining the annual variation in this ratio, despite the disruptive impact of the COVID-19 pandemic, there hasn't been a significant surge in the overall volume of research. This observation suggests a sustained commitment to this field over time. Furthermore, the diverse range of topics outlined in Table 2 underscores the urgent need for active engagement and contribution from the Korean academic community. This highlights not only the volume but also the breadth of approaches required to make meaningful strides in this field.

The volume of research specified for each Topic Category in Table 3, provides insights into where scholarly interests are concentrated. The total number and proportion of studies demonstrate a relatively uniform distribution ranging from 10% to 30%, with the exceptions of categories 3 and 8. Category 3 stands out with an overwhelming proportion of 60.2%, indicating significant interest in meso-level healthcare systems, particularly at the level of healthcare institutions.

Category 8, 'Psychiatric Care and Mental Health', exhibits a minimal proportion of 3.7%, indicating a lower volume of system engineering approaches within this field. This suggests that within the realm of mental health, an area of significant societal importance, there remains untapped potential for system engineering researchers. The independent establishment of mental health care as a category is particularly intriguing. It encompasses significant issues relevant to South Korea, such as depression treatment, while also addressing critical concerns in the United States, such as opioid addiction. Given the significance of psychiatric disorders in South Korea, evidenced by metrics like the highest suicide rate among OECD countries, domestic research addressing our nation's challenges in this area would be highly beneficial.

By looking deeper into the yearly change of this volume, several observations could be explained by associating with the characteristics of each category, as well as external factors. Healthcare Resource Management (Category 3), was the only one labeled as a 'well-established topic'. The substantial proportion of Category 3 within the field, coupled with a recent decline in interest, suggests that the research community's initial focus was on resource allocation issues. Such focus aligns with typical and prevalent problems encountered in industrial and management engineering. As significant progress has been made in addressing this problem, other more complex and unresolved issues within the system have emerged.

In contrast, the category of 'emerging topics' encompasses areas such as Public Health and Epidemiology (Category 4), Health Informatics and AI (Category 10), and Social Determinants of Health and Health Inequalities (Category 11). These increases can be attributed to external factors such as pressing societal issues and advancements in technology. Category 4 has been heavily influenced by the global impact of the COVID-19 pandemic, becoming a strong mainstream of research interest since late 2019. Category 10 demonstrates increasing attention and efforts to integrate advanced AI into various components of healthcare services. Finally, category 11 reflects the rising societal awareness and interest in equity and fairness. Addressing basic human welfare from the perspective of marginalized groups' rights and political correctness is indeed not only a trend but also a fundamental aspect of this academic field.

A closer examination of the sub-topics in Appendix 7 suggests that this field deals with decision-making across various dimensions. These range from individual healthcare treatment decisions to institutional decisions regarding the treatment and diagnosis of multiple patients' health conditions, and even national-level policy decisions aimed at monitoring public health. The diverse nature of these decision-making processes highlights the breadth of issues that can be addressed within the realm of industrial engineering. Additionally, alongside approaches aimed at enhancing system efficiency, numerous studies focus on integrating new technologies into healthcare systems and evaluating their feasibility and efficacy.

Ultimately, Korea's Healthcare Systems Engineering research must remain adaptive and innovative to address evolving challenges. collaboration, both domestically and internationally, is crucial for making impactful contributions to healthcare advancements. By leveraging interdisciplinary approaches and global insights, Korea can drive positive change in healthcare delivery, ultimately improving patient outcomes and societal well-being.

## 7.2 Discussion on the review process

In the realm of academic review, the methodological application of ChatGPT has unveiled both strengths and limitations. The largest advantage of employing ChatGPT lies in its capability to 'read' the information efficiently, requiring minimal time and human resources. The execution time for the entire process, encompassing the reading and organization of approximately 40,000 research abstracts, was completed in less than 3 weeks, including all trial and error phases. With a well-refined process, it is anticipated that the pure execution could be accomplished in less than a week.

This efficiency significantly lowers the barrier of navigating a research field, thereby facilitating quick understanding, especially for those newcomers to the domain. In the rapid emergence and evolution of numerous research fields, this analytical technique holds the potential to address the absence of experienced researchers in such fields. Moreover, the categorization of topics derived from a somewhat unorganized landscape of studies can be applied to various tasks requiring the structuring of a research community. For instance, it can facilitate the organization of academic conference sessions and timetables by grouping similar studies.

Furthermore, the accessibility of ChatGPT utilization represents another significant advantage. In contrast to conventional topic mining approaches requiring specific NLP models for each process component (Keyword Extraction, Categorization,

Classification, and Summarization), ChatGPT's implementation enables researchers, regardless of their expertise in NLP, to customize these processes simply by writing their requirements in sentences. This accessibility lowers the implementation barriers, enabling researchers to acquire reviews precisely tailored to their needs of insight.

Despite these notable advantages, there are several limitations of this review framework. The generative nature of the GPT model raises concerns regarding the reliability and consistency of its responses. Although pilot experiments were conducted to assess the reliability and consistency, they remain uncertain in another dataset. The fact that the derived Topic Categories lack of mutual exclusiveness also raises concern. As the Topic Categorization process and the Categorization process were executed serially, the derived categories are more like a summary of the dataset, in the format of a list, which does not guarantee mutual exclusiveness of the Classification process.

Furthermore, there exist limitations in explaining the result of the analysis, such as relating them with external societal factors, The necessity of human researchers' curation on the product of the Topic categorization process also remains as a limitation, as it keeps our framework from being fully automatic. The model's inability to reference specific instances of studies also poses a challenge. While indexing all studies and prompting the model to specify instances within a given context could address this issue, it remains susceptible to the consistency challenges of ChatGPT. Overall, ChatGPT-generated reviews may not fully encapsulate the depth and breadth of information compared to handmade academic reviews.

The shortage of this analysis also arises from the format of the data. Given that the dataset was in the form of abstract paragraphs of the conference presentations, the containing information may not be sufficiently comprehensive. Many abstract paragraphs focus primarily on defining their research question and providing their findings, they often lack details on the methodology. This limitation could be addressed by utilizing full manuscripts as the data source and modifying the framework to classify the methods along with the problem domains of the studies. Such an approach would become possible as the input token limit of ChatGPT continues to expand.

Despite the significant advantages offered by this review framework, several limitations need to be acknowledged. The generative nature of the GPT model raises concerns regarding the reliability and consistency of its responses. While pilot experiments were conducted to assess these issues, uncertainties remain when applied to different datasets. Furthermore, the lack of mutual exclusiveness in the derived Topic Categories is a notable concern. The product of the Topic Categorization process resembles a summary of the dataset, presented in a list format. This does not guarantee the following Classification process to label each entry to a single category. Additionally, the necessity for human researchers' curation in the Topic categorization process remains a limitation, preventing the framework from being fully automatic.

Moreover, the model's inability to reference specific instances of studies poses a challenge. While indexing all studies and prompting the model to specify instances within a given context could address this issue, it remains susceptible to consistency challenges inherent in ChatGPT. Overall, ChatGPT-generated reviews may not fully capture the depth and breadth of information compared to manually crafted academic reviews.

We recognize that our product may not reach the depth of handcrafted reviews by veteran researchers, as it lacks some of the critical components of regular academic reviews (e.g., evaluations, criticisms, and citations). However, our endeavor significantly lowers the barrier of deriving the structure of research topics, that can function as an accessible guidebook of the academic field. Furthermore, we hope that our work serves as a proposal to facilitate the easy production of necessary reviews in other fields as well.

## 8. Conclusion

In summary, we present an efficient academic review framework leveraging ChatGPT across each analysis process component. Our comprehensive review of the Healthcare Systems Engineering field comprises 11 topic categories, providing a descriptive backbone of the landscape of research. Following this categorization, we conduct quantitative trend analysis and in-depth sub-topics within each category. As shown in the products of the review, the application of industrial engineering methodologies in healthcare is not only prevalent but also crucial for addressing complex systematic problems.

Our analysis reveals the urgent need for Korea's academic community to engage in healthcare systems engineering research, aligning with global imperatives and societal demands. The areas highlighted in this review—such as public health, mental healthcare, application of AI, and social determinants of health—represent not only emerging research efforts but also opportunities for pioneering work.

The utilization of ChatGPT in academic reviews offers advantages in terms of rapid information acquisition and accessibility. Although concerns regarding reliability and consistency remain, the rapid evolution of the ChatGPT model holds promise for enhanced review quality in the future. We hope this review serves as a valuable resource for researchers in the Healthcare Systems Engineering domain and beyond.

# Appendix

## Appendix 1 Prompt for the Filtering Process

[title of a study]

[abstract paragraph of a study]

Above is a research title and an abstract.

Please indicate whether the research problem is directly related to the healthcare system, hospital operation, medical care, emergency system, patient, or public health.

Your answer should be either 'Yes' or 'No'.

## Appendix 2 Prompt for the Keyword Extraction Process

[title of a study]

[abstract paragraph of a study]

The text provided above consists of a research title and an abstract. Please identify the research keywords from this abstract, which is part of a presentation at a conference.

The conference focuses on the application of Operations Research techniques and Management Science methods to address issues within the Healthcare System Domain.

Please provide 5 keywords related to the healthcare system domain of the research problem, and 5 keywords related to the system engineering methodology or management technique.

Each keyword may consist of 1 to 3 words, and your response should be a comma-separated list of keywords. Please aim for the keywords to be as mutually exclusive as possible.

**Appendix 3 Prompt for the Topic Categorization Process**

[list of extracted keywords]

I would like to find out several main categories of Industrial/System Engineering and Management Science in the Healthcare system domain.

Please observe the provided list of keywords, and try merging and allocating those that are similar to each other.

Please provide them in a numbered list, with some detailed descriptions about each category.

**Appendix 4 Examples of raw outputs of topic categorization process**

| Output 1 |
|---|
| 1. Healthcare Operations Management |
| 2. Chronic and Communicable Diseases |
| 3. Health Policy and Systems |
| 4. Health Technology and Data Analysis |
| 5. Patient Care Management |
| 6. Mental Health Services |
| 7. Emergency Services and Critical Care |
| 8. Preventive Healthcare and Wellness |
| 9. Healthcare Economics and Public Health |
| 10. Epidemiology and Disease Transmission |

| Output 2 |
|---|
| 1. Health Services and Clinical Practices |
| 2. Disease Management and Treatment |
| 3. AI in Healthcare |
| 4. Patient Experience and Satisfaction |
| 5. Health Policies |
| 6. Advanced Healthcare Supplies and Facilities |
| 7. Healthcare Workforce Management |
| 8. Healthcare Innovation and Transformation |
| 9. Health and Society |
| 10. Hospital Operations Management |

| Output 3 |
|---|
| 1. Disease Monitoring and Public Health |
| 2. Hospital Service and Patient Care |
| 3. Healthcare Policies and Forecasting |
| 4. Pharmaceuticals and Medical Supplies |
| 5. Machine Learning and Data Science in Healthcare |
| 6. Treatment Decision Making |
| 7. Social Determinants and Health Disparities |
| 8. Health Systems and Health Economics |
| 9. Mental Health and Substance Use |
| 10. Health Infrastructure and Facilities Management |
| 11. Use of Technologies for Healthcare Improvement |
| 12. Lifestyle and Dietary Factors in Health |

## Appendix 5 Prompt for the Classification Process

[title of a study]

[abstract paragraph of a study]

Above is a research abstract. Please summarize it in one sentence.

After that, please guess which of the following research topic categories it belongs to.

[numbered list and explanations of 11 topic categories]

Your answer should both include one sentence summary, and also the number of assigned topic category. I want the summary sentence in the {braces}, and the numbers to be in a [square brackets].

You may choose multiple categories if needed. If none of the categories can properly describe the research, please answer the topic category as 0.

Please think carefully.

Answer example:

{Summary Sentence blah blah blah}

topic category = [3, 7]

**Appendix 6 Prompt for the sub-topic mining Process**

[name of a specific topic category]

[explanation of the respective topic category]

Above is one of the main categories of research topics of Healthcare systems engineering.

And followings are one-sentence-summaries of each research that belongs to the above topic category.

Give me 4~5 sub-categories that best describe the mainstream of the above topic category, based on the following research summaries. I want a bulleted list with some explanations.

[list of one-sentence-summaries]

**Appendix 7 Extracted Sub-topics for each topic categories**

1. Clinical Decision Making
    (1) Personalized Treatment Strategy
        - Crafting and Fine-tuning Individualized Treatment Plans
        - Program Development for Cancer, Diabetes, and Persistent Disease Treatment
        - Machine Learning from Clinical Data
        - Mathematical Modeling for Therapeutic Optimization
    (2) Diagnostic and Screening Policy
        - Creating Diagnostic and Testing Protocols Using Health Records and Demographics
        - Enhancing Diagnostic Precision and Decision Support Tools
    (3) Predictive Model
        - Modeling Patient Trajectories and Prognoses
        - Systems for Supporting Clinical Decisions
    (4) Risk Assessment
        - Identifying Patient Groups at Elevated Risk and Analyzing Mortality Data
        - Assessing Dangers Throughout the Treatment Journey

2. Clinical Decision Making

   (1) Home Care and Remote Monitoring

      - Evaluating the Effects, Efficiency, and Utilization of Remote Monitoring Implementation

      - Quantitative Validity Testing of Home Care (Including Patient Medical Accessibility)

   (2) Choice Model and Preference Analysis

      - Analysis of Patient Preferences and Choice Factors for Healthcare Services/Institutions

      - Data-Driven Research, and Survey-Based Studies

   (3) Patient Satisfaction and Experience

      - Patient Satisfaction and Experience Improvement Strategies

      - Patient flow, Care pathway Optimization

      - Patient-Centered Care, Accessibility, Service Quality

   (4) Patient Engagement

      - Patient Involvement in Treatment Decision-Making

      - Technology for Providing Patient Information

3. Healthcare Resource Management

   (1) Resource Allocation

      - Institute, Care Facility Allocation

      - Medical Supplies, Human Resource, Physician Burnout

      - Capacity Planning

   (2) Patient Scheduling

      - Outpatient Appointment Scheduling

      - Waiting Time Reduction

   (3) Healthcare Supply Chain Management

      - Applying SCM Techniques to Healthcare Systems

      - Inventory Management, Logistics, Tracking

   (4) Analytics and Decision Support

      - Data-Driven/Optimization-Based Decision-Making Models in Hospital

      - Patient Admission Decision, Hospital/ED Operation

4. Public Health and Epidemiology

(1) Disease Modeling

- Mathematical Modeling/Simulation on Disease Spread
- Disease Dynamics Prediction Model

(2) Resource Allocation

- Resource Allocation Under the Surge Demand in Epidemic
- Screening, Vaccine, Care Units

(3) Public Health Intervention

- Validation Research on the Efficiency and Effectiveness of Infection Control Strategies
- Pharmaceutical/Non-Pharmaceutical Interventions

(4) Behavioral Operation Research

- Policy/Intervention Design Incorporating Patient's/Population's Behavior
- Epidemics, Chronic Disease, Patient's Choices

5. Health Policy and Economics

    (1) Payment System and Health Insurance

    - Medicare/Medicaid
    - Mathematical Models on Payment Systems
    - Fee-for-Service, Bundled Payment, etc.

    (2) Healthcare Economics

    - Cost-Effectiveness of Healthcare Services
    - Economic Evaluation on the Operation Strategies

    (3) Public/Private Hospitals

    - Comparative Analysis on Public/Private Medical Services
    - Equilibrium Points between Services
    - Supportive Policies, Subsidization

    (4) Care Organization

    - Accountable Care Organization (ACO)
    - Relationships between Healthcare Institutions
    - Collaboration/Competition

6. Preventive Healthcare and Well-Being

    (1) Preventive Treatments

    - Vaccination Program Design and Assessment
    - Disease Screening, Disease Damage Control

(2) Personal Behavior

- Analysis of the Influence of Dietary Habits, Exercise, Lifestyle, etc.

(3) Public Health Promotion

- Health Promotions, Education Programs
- Health-Related Subsidy Program

(4) Safety Management

- Risk Management, Safety Engineering
- In-Hospital Falls, Medical Accidents, etc.

7. Emergency and Critical Care

   (1) Emergency Department Operations

   - ED Operating Strategy Design/Optimization
   - Patient Triage
   - Waiting Time Reduction, Patient Flow

   (2) Disaster Response

   - Resource Management Strategies under Disaster Situations/Surge Demand
   - Decision Support System for Disaster Response

   (3) Ambulance and EMS

   - Ambulance Operating Strategy/Location Allocation

   (4) ICU Operations

   - ICU Operating Strategy
   - ICU Resource Management
   - Patient Admission Decision Support System

8. Psychiatric and Mental Health

   (1) Depression Treatment

   - Personalized Therapy Design Incorporating Patients' Medication Non-Adherence
   - Depression Prediction Model Using EHR Data

   (2) Addiction Management

   - Substance Abuse, Alcoholism
   - Opioid Use Disorder

   (3) Screening and Detection

   - PTSD, Epilepsy, Parkinson's
   - Risk Assessment, Screening Strategy

(4) Social Determinants on Mental Health
      - Impact of Social Media, Celebrity Suicides
      - Setting Gatekeepers in Social Media for Suicidal Prevention

9. Innovations in Healthcare
   (1) Telemedicine
      - Telemedicine, E-consult, Online Patient Portals
      - Adoption Effect, Utilization Strategies
   (2) Electronic Health Record
      - Data Collection, Database Construction
      - Wearable Devices, IoT
      - Remote Monitoring
      - Clinical Decision Support System

10. Health Informatics and AI
    (1) Clinical Decision Support System
       - EHR, Clinical Data, Medical Imaging AI
       - Diagnosis Supporting System
       - Patient Monitoring
    (2) Predictive Model
       - Patient Progress and Risk Prediction
       - Demand Prediction on Hospital/Regional Scale
    (3) Hospital Operation Support
       - Efficiency in Healthcare Cost and Resource Management
       - Outpatient Scheduling

11. Social Determinants of Health and Health Inequalities
    (1) Health Inequalities
       - Socioeconomic, Racial, Environmental
       - Measuring Health Disparities and Healthcare Accessibility Inequalities
       - Designing Assistance and Subsidy Programs
    (2) Maternal and Childhood Healthcare
       - Maternal Mortality Reduction/Policy Development
       - Healthcare Delivery System Model for Improving Neonatal and Infant

(3) Underserved Communities

- Designing Public Health Policies for Medically Underserved Populations

- Refugees, Homeless, Mental Health Patients, etc.

- Developing Low-Cost Healthcare Service Models

(4) Global Health Disparities

- International Disparities in Healthcare Services and System Levels and Solutions

- International Infectious Disease Response and Optimization of Vaccine Procurement Strategies

# References

Alshami, A., Elsayed, M., Ali, E., Eltoukhy, A. E., & Zayed, T. (2023). Harnessing the power of ChatGPT for automating systematic review process: Methodology, case study, limitations, and future directions. *Systems*, 11(7), 351.

Chung, J. J. Y., Kamar, E., & Amershi, S. (2023). Increasing diversity while maintaining accuracy: Text data generation with large language models and human interventions. *arXiv preprint arXiv:2306.04140*.

Clavié, B., Ciceu, A., Naylor, F., Soulié, G., & Brightwell, T. (2023, June). Large language models in the workplace: A case study on prompt engineering for job type classification. *International Conference on Applications of Natural Language to Information Systems* (pp. 3-17). Cham: Springer Nature Switzerland.

cOASIS. (2017). INFORMS Annual Meeting 2017 Itinerary Builder. https://www.abstractsonline.com/pp8/#!/4471. Accessed October 2023.

cOASIS. (2018). INFORMS Annual Meeting 2018 Itinerary Builder https://www.abstractsonline.com/pp8/#!/4701. Accessed October 2023.

cOASIS. (2019). INFORMS Annual Meeting 2019 Itinerary Builder. https://www.abstractsonline.com/pp8/#!/6818. Accessed October 2023.

cOASIS. (2019). INFORMS Healthcare 2019 Conference Itinerary Builder. https://www.abstractsonline.com/pp8/#!/6830. Accessed October 2023.

cOASIS. (2020). INFORMS Annual Meeting 2020 Itinerary Builder. https://www.abstractsonline.com/pp8/#!/9022. Accessed October 2023.

cOASIS. (2021). INFORMS Annual Meeting 2021 Itinerary Builder. https://www.abstractsonline.com/pp8/#!/10390. Accessed October 2023.

cOASIS. (2021). INFORMS Healthcare 2021 Conference Itinerary Builder. https://www.abstractsonline.com/pp8/#!/9318. Accessed October 2023.

cOASIS. (2022). INFORMS Annual Meeting 2022 Itinerary Builder. https://www.abstractsonline.com/pp8/#!/10693. Accessed October 2023.

cOASIS. (2023). INFORMS Healthcare 2023 Conference Itinerary Builder. https://www.abstractsonline.com/pp8/#!/10855. Accessed October 2023.

Erren, T. C., Cullen, P., & Erren, M. (2009). How to surf today's information tsunami: on the craft of effective reading. *Medical hypotheses*, 73(3), 278-279.